# Nanostructured Ceramic Oxides with a Slow Crack Growth Resistance Close to Covalent Materials

Jérôme Chevalier[1], Sylvain Deville[1], Gilbert Fantozzi[1], José F. Bartolomé[2], Carlos Pecharroman[2], José S. Moya[2], Luis A. Diaz[3], and Ramon Torrecillas[3]

[1] Materials Department (UMR CNRS 5510)
National Institute for Applied Sciences
20 Avenue Albert Einstein, 69621 Villeurbanne Cedex, France

[2] ICMM, Spanish Research Council (CSIC)
Campus Cantoblanco, Madrid, Spain

[3] INCAR, Spanish Research Council (CSIC)
La Corredoria s/n Ap.73, 33080 Oviedo, Spain

## Abstract

Oxide ceramics are sensitive to slow crack growth because adsorption of water can take place at the crack tip, leading to a strong decrease of the surface energy in humid (or air) conditions. This is a major drawback concerning demanding, long-term applications such as orthopaedic implants. Here we show that a specific nanostructuration of ceramic oxides can lead to a crack resistance never reached before, similar to that of covalent ceramics.

Ceramic engineers are very often confronted with a dilemma when considering ceramic materials for structural applications. Oxide ceramics, such as alumina or zirconia, are quite easy to process (sintering under normal, air conditions at relatively moderate temperatures <1600°C) but are prone to slow crack growth (SCG) if used under water, humid, or body fluid conditions. Conversely, covalent ceramics, such as silicon carbide or silicon nitride, are considered to be almost insensitive to SCG but require much more effort during processing (high sintering temperature >1700°C under inert atmosphere). Oxide ceramics are sensitive to SCG because adsorption of water can take place at the crack tip, leading to a strong decrease of the surface energy in humid (or air) conditions. This process, first reported and modeled for a glass,[1] is now accepted for all oxide ceramics.[2] This is a major drawback concerning demanding, long-term applications, such as orthopaedic implants. Here we show that a specific nanostructuration of ceramic oxides can lead to a SCG resistance never reached before, similar to that of covalent ceramics.

As a general trend, the susceptibility of ceramics to SCG is discussed on the basis of a $V$ (crack velocity) versus $K_I$ (stress intensity factor) diagram *($K_I$ representing the stresses at the tip of a crack or any preexisting defect such as a pore, a scratch, etc, in the ceramic).* Recently, the presence of a threshold in the stress intensity factor, under which no crack propagation occurs, has been the subject of important research in the ceramic field.[3] This threshold corresponds to a crack equilibrium with a null crack velocity. For ceramic joint prostheses, for example, this threshold, $K_{I0}$, determines a safe range of use. The comparative sensitivity of ceramics to SCG can be plotted in a normalized $V-K_I/K_{IC}$ diagram, where $K_{IC}$ is the toughness. The higher the slope of the diagram, i.e., the higher the $K_{I0}/K_{IC}$ ratio, the lower the sensitivity to SCG by stress assisted corrosion. Figure 1 represents a schematic summary of results obtained in different ceramics under similar conditions (i.e., double torsion).[4] These results were obtained on ceramic materials without extrinsic reinforcement mechanisms (small grain size, less than 2 μm), without sintering additives in the case of covalent materials, avoiding the influence of glassy phase. The results illustrate the commonly accepted idea that the higher the covalent to ionic bonding ratio, the lower the susceptibility to SCG. This is directly related to the atomic structure of the material. Since SCG is one of the major issues in the ceramic community, there has been a continuous effort to control the SCG in ceramics by tailoring their microstructure. In the 80's particularly, there was a trend to increase extrinsic reinforcement mechanisms such as crack bridging (i.e., by increasing the grain size and/or changing the grain morphology) to increase crack propagation resistance.[5] This was demonstrated to be effective on SCG resistance for long cracks.[6] However, it is clear today that this mechanism is far from being totally effective for real, short cracks.[7] Moreover, these extrinsic mechanisms are degraded partially, if not totally, under cyclic fatigue conditions.[8]

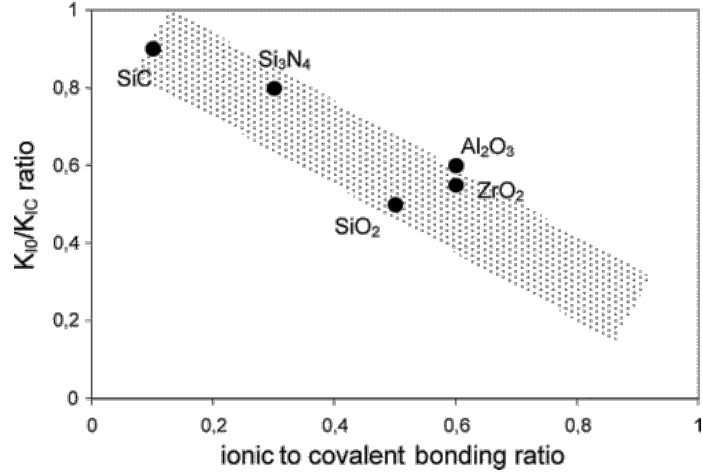

Figure 1: $K_{I0}/K_{IC}$ ratio versus ionic/covalent ratio for different ceramics.

In the present investigation, we shift to a different paradigm. Instead of extrinsic we propose a more intrinsic reinforcement mechanism, as it is the development of a compressive residual stress field into the composite matrix that can be applied to real cracks under any loading condition. This is the case of an alumina$-n$ZrO$_2$ composite, where zirconia nanoparticles are evenly distributed into the alumina matrix mainly at intragranular positions. If the particles are in tetragonal symmetry, as a consequence of the thermal expansion mismatch ($\Delta\alpha \approx 4 \times 10^{-6}$ K$^{-1}$), they must be subject to a very high isostatic tensile stress ($\sim$1 GPa) inducing, according to the Selsing[9] expression, a compressive residual stress into the continuous alumina matrix.

To achieve this goal, a modified colloidal route was conducted to synthesize the nanostructured composite powder. This processing route was described in detail elsewhere[10] and already applied to composites with improved mechanical properties.[11] It consists of doping a stable suspension of a high-purity alumina powder (Condea HPA 0.5, with an average particle size of 0.45 µm and a surface area of 10 m²/g) in ethanol absolute (99,97%) by dropwise addition of a diluted (2/3 vol % Zr alkoxide, 1/3 vol % ethanol absolute) zirconium alkoxide (Aldrich zirconium IV propoxide, 70 wt % solution in 1-propanol). In the present work, a low amount of zirconia precursor was added, to obtain composites with only 1.7 vol. % (2.5 wt %) zirconia nanoparticles. After drying under magnetic stirring at 70 °C, the powders were thermally treated at 850 °C for 2 h in order to remove organic residues and were subsequently attrition milled with alumina balls for 1 h. Green compacts were then obtained by a traditional slip casting method. The optimum sintering to obtain the desired nanostructural distribution of zirconia particles consisted of a thermal treatment of 1600 °C for 2 h. Figure 2 shows the microstructure of the material, consisting in zirconia nanoparticles evenly distributed in the alumina matrix. Those zirconia particles were found to be mainly (>70%) intragranular, with almost perfect spherical shape ($D_{50} \approx 150$ nm). These particles are well below the critical size for phase transformation.[12] The alumina grains ($D_{50} \approx 5$ μm) exhibit a relatively broad grain size distribution, and many grains are tabular or elongated.

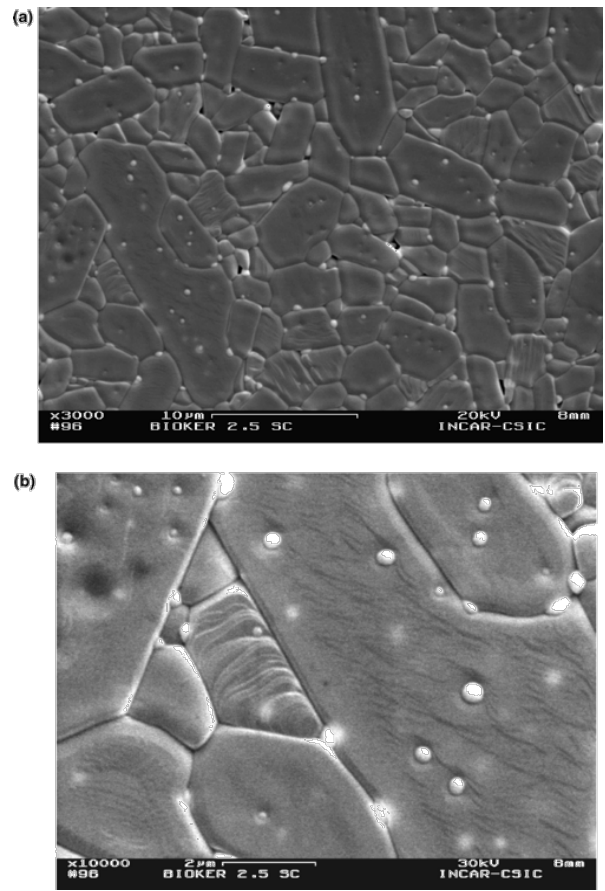

Figure 2: (a) Scanning electron microscopy image of the nanostructured composite. (b) Close-up showing nanosized zirconia particles embedded in the alumina matrix.

Short-crack resistance curves (Figure 3) were measured and analyzed by the indentation−strength in bending (ISB) method.[13] The behavior of the present nanostructured composite was compared to that of alumina ceramic with a similar microstructure and grain size distribution.[14] For monolithic $Al_2O_3$, the initial toughness is about 3 MPa m$^{1/2}$ and rises to a maximum value of 5 MPa m$^{1/2}$ within a 1000 μm extension, because of grain bridging by large elongated $Al_2O_3$ grains behind the crack tip.[15] Whereas the toughness of the alumina is enhanced in the long-crack region, it is low in the short-crack region. Moreover, this so-called "R-curve" effect due to extrinsic mechanisms can be degraded under specific cyclic conditions.[8] Therefore, in applications where small cracks are expected and under alternative loading, for example in bearing applications and for other well-polished surface finishes, as it is the case of total joint (hip and knee) replacement, R-curve materials may not be desirable. In comparison, the nanostructured composite exhibits a peculiar behavior, with a constant toughness value of 6 MPa m$^{1/2}$, i.e., twice that of the short-crack resistance of pure alumina. This is a major progress, since we have to keep in mind that only 1.7 vol % of zirconia is present in the composite.

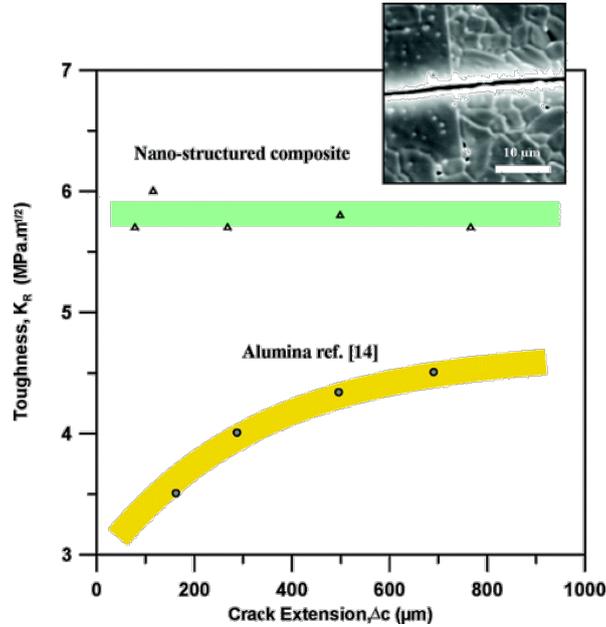

Figure 3: Short-crack resistance curve of the present nanostructured composite (top), compared with an alumina ceramic with similar microstructure (ref 14). A SEM micrograph of the nanocomposite showing an indentation crack without grain bridging is inserted.

To confirm the strong potential of the present nanocomposite, $V-K_I$ curves were conducted by the double torsion method, i.e., the same method that was used for the generation of the results summarized in Figure 1. The results are given in Figure 4 and compared to 2 µm grain size alumina (without crack bridging) and 0.5 µm grain size zirconia, which are considered as the standard technical oxide ceramics. Again, the nanocomposite exhibits a peculiar behavior, with a slope of the V$-K_I/K_{IC}$ diagram and a $K_{I0}/K_{IC}$ ratio close to covalent ceramics! This is illustrated in Figure 5, where the present nanostructured composite is compared to ceramics with different covalent-to-ionic bonding ratios. In this diagram, the Al$_2$O$_3-n$ZrO$_2$ lies between Si$_3$N$_4$ and SiC.

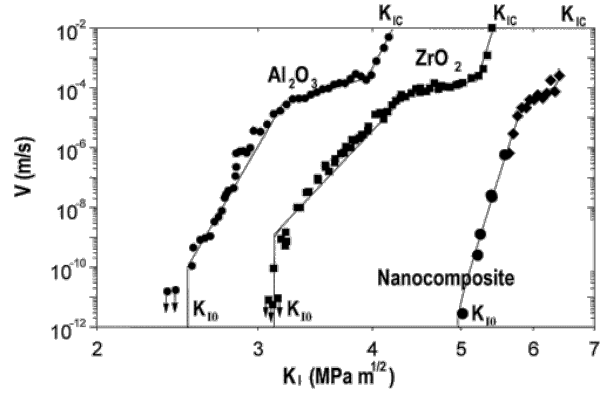

Figure 4: $V-K_{\mathrm{I}}$ curve of the present nanostructured composite, compared with standard alumina and zirconia.

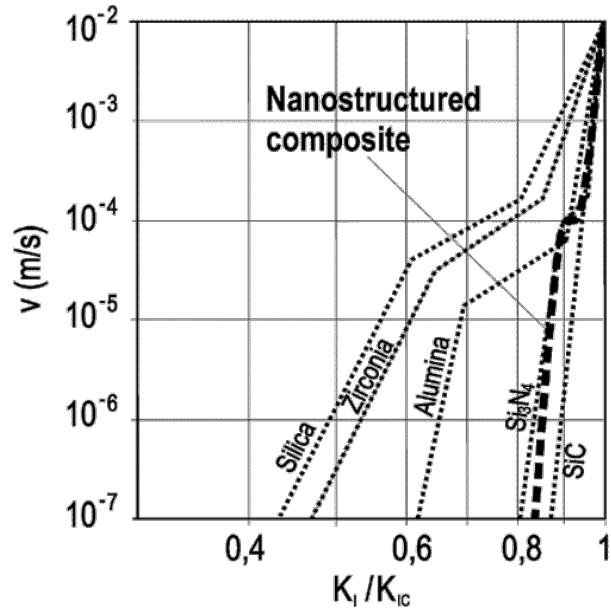

Figure 5: $V-K_{\mathrm{I}}/K_{\mathrm{IC}}$ curve of the present nanostructured composite, compared with ionic and covalent ceramics, illustrating its peculiar behavior versus slow crack growth.

This dramatic increase of crack growth resistance cannot be attributed to crack bridging, since the R-curve was proven to be flat and the mode of failure predominantly transgranular (Figure 3).

To determine the presence of a residual stress field in the alumina matrix, high angular precision diffraction patterns (monocromatized incident beam Philips Xpert diffractometer) were recorded between 27 and 45 degrees in pure sintered alumina and alumina/nanozirconia (1.7 vol %) composite plates, both obtained by slip casting. Small angular displacements were found in the $\alpha$-alumina peaks of nanocomposite plate corresponding to the following planes: (104), (110), (006), (113). According to these data, compressive strains of $3 \times 10^{-4}$ and $2 \times 10^{-4}$ were found for the $a$ and $c$ axes of the $\alpha$-alumina matrix. These strains correspond to a compressive average stress of $150 \pm 50$

MPa. It should be noted that, according to the Selsing expression for thermal stresses (taking into account the XRD pattern of the nanocomposite plate, the zirconia grains were considered to be in tetragonal symmetry, no sign of monoclinic zirconia was detected), zirconia grains must hold an isostatic tensile stress of about 1.4 GPa. Additionally, the effect of volume concentration slightly raises the stress through

$$p_c = \frac{\Delta\alpha\Delta T}{\dfrac{1 - 2v_c}{E_c} + \dfrac{1 + v_s + 2f(1 - 2v_s)}{2E_s(1 - f)}}$$

(eq. 1)

where $E_c$, $v_c$, $E_s$, and $v_s$ are, respectively, the Young moduli and Poisson coefficients for included particles and matrix; $f$ is the volume concentration of included particles defined by:

$$f = \left(\frac{R_1}{R_2}\right)^3$$

(eq. 2)

In this expression, $R_1$ refers to the particle radius, and $2R_2$ is the average distance between particles. However, the stresses in the alumina matrix decay from the zirconia/alumina interface following a $r^3$ law, modified for low concentrations. In this sense, a spatial average must be carried out:

$$\langle\sigma\rangle = \frac{\int \sigma(\vec{r})dV}{V} = 3\frac{\int_{R_1}^{R_2} \sigma(r)r^2 dr}{R_2^3 - R_1^3}$$

(eq. 3)

Solving, we get the average Von-Mises stress for a sphere:

$$\langle\sigma_{VM}\rangle = -\frac{3}{2}\frac{p_c f}{1 - f}\ln\,(f)$$

(eq. 4)

It should be noted that eqs 1−4 are valid only for low inclusion concentrations (<3vol %) because they have been derived assuming there is no overlapping between the stress fields associated to the inclusions. For larger amounts, the dependence of the $\sigma_{VM}$ vs $f$ reaches a maximum for a specific concentration range, which mostly depends on the microstructure of the composite, then subsequently decreases when $f$ tends to 1. Substituting values for 1.7 vol. % zirconia spheres included in an alumina matrix, we get an approximated valued of 100 MPa for this stress. This compressive residual stress field is consistent with the obtained experimental value ~150 MPa. This value is more than twice the one determined in the case of a mullite−alumina laminate (≈60 MPa) that was able to arrest the fatigue cracks and to improve the fatigue life of the material.[16] In the nanocomposite, this compressive residual stress field retards the crack front opening as an external stress is applied, to stand in competition with the stress corrosion. The relative effect of the residual stress intensity factor is higher for lower applied stress intensity factors. Because of this, a higher slope in the nanocomposite $V−K_I/K_{IC}$ curve is obtained, which is similar to the ones observed in covalent materials (Figure 5). The engineering implication of this increase of the $V−K_I/K_{IC}$ curve slope is a lower sensitivity to delayed failure, as slow crack growth occurs for larger applied $K_I/K_{IC}$ values.

In summary, here we show that the compressive residual stress field developed because of the presence of a small volume fraction of evenly distributed zirconia nanoparticles is responsible for the drastic change in the overall resistance to slow crack growth of the alumina−zirconia nanocomposite. This result opens a new avenue of developing oxide ceramic-based nanostructured composites for structural applications since they offer crack resistance similar to covalent materials without their major drawbacks associated with processing and machining.

## Acknowledgments

This work was supported by EU under projects reference BIOKER GRD2-2000-25039 and IP-NANOKER FP6-515784-2, and by the Spanish Ministry of Education and Science under project reference MAT2003-04199-C02. J.F.B. has been supported by Ministry of Science and Technology and CSIC under the "Ramón y Cajal" Program cofinanced by the European Social Fund.

## References

1. Michalske, T. A.; Freiman, S. E. *J. Am. Ceram. Soc.* **1983**, *66*, 284−288.
2. Lawn, B. *Fracture of Brittle Solids*; Cambridge University Press: Cambridge, 1993
3. Wan, K. T.; Lathabai, S.; Lawn, B. R. *J. Eur. Ceram. Soc.* **1990**, *6*, 259−268.
4. Chevalier, J. *Ceramics for biomedical applications* (in French), Habilitation à Diriger des recherches; Edited by INSA de Lyon, France, 96 pages, 2001
5. Swanson, P. L.; Fairbanks, C. J.; Lawn, B. R.; Mai, Y. W.; Hokey, B. J. *J. Am. Ceram.* Soc. **1987**, *70*, 279−289.
6. Lathabai, S.; Lawn, B. R. *J. Mater. Sci.* **1989**, *24*, 4298−4306.
7. Fett, T.; Munz, D. *J. Mater. Sci. Lett.* **1992**, *11*, 257−260.

8. El Attaoui, H.; Saadaoui, M.; Chevalier, J.; Fantozzi, G. *J. Am. Ceram. Soc.*, in press 2005.

9. Selsing, J. *J. Am. Ceram. Soc.* **1961**, *44*, 419−426.

10. Schehl, M.; Díaz, L. A.; Torrecillas, R. Alumina nanocomposites from powder-alcoxide mixtures, *Acta Mater.* **2002**, *50*, 1125−1139.

11. Chevalier, J.; De Aza, A. H.; Schehl, M.; Torrecillas, R.; Fantozzi, G. *Adv. Mater.* **2000**, *12*, 1619−1622.

12. Heuer, A. H.; Claussen, N.; Kriven, W.; Rühle, M. *J. Am. Ceram. Soc.* **1982**, *65*, 642−650.

13. Bartolomé, J. F.; Diaz, M.; Moya, J. S. *J. Am. Ceram. Soc.* **2002**, *85*, 2778−2784.

14. Kovar, D.; Bennison, J.; Readey, J. *Acta Mater.* **2000**, *48*, 565−578.

15. Vekinis, G.; Ashby, M. F.; Beaumont, P. W. R. *Acta Metall. Mater.* **1990**, *38*, 1151−1162.

16. Bartolomé, J. F.; Moya, J. S.; Requena, J.; Llorca, J.; Anglada, M. *J. Am. Ceram. Soc.* **1998**, *81*, 1502−1508.